\documentclass{elsart5p}

\usepackage{amsmath,graphicx}

\begin{document}

\begin{frontmatter}

\title{Unraveling gravity beyond Einstein with \\ extended test bodies}

\author{Dirk Puetzfeld\corauthref{cor1}}
\address[cor1]{ZARM, University of Bremen, Am Fallturm, 28359 Bremen, Germany}
\ead{dirk.puetzfeld@zarm.uni-bremen.de}
\ead[url]{http://puetzfeld.org}

\author{Yuri N.\ Obukhov\corauthref{cor2}}
\address[cor2]{Theoretical Physics Laboratory, Nuclear Safety
  Institute, Russian Academy of Sciences, B.\ Tulskaya 52, 115191
  Moscow, Russia} 
\ead{yo@thp.uni-koeln.de}

\begin{abstract}
The motion of test bodies in gravity is tightly linked to the conservation laws. This well-known fact in the context of General Relativity is also valid for gravitational theories which go beyond Einstein's theory. Here we derive the equations of motion for test bodies for a very large class of gravitational theories with a general nonminimal coupling to matter. These equations form the basis for future systematic tests of alternative gravity theories. Our treatment is covariant and generalizes the classic Mathisson-Papapetrou-Dixon result for spinning (extended) test bodies. The equations of motion for structureless test bodies turn out to be surprisingly simple, despite the very general nature of the theories considered. 
\end{abstract}

\end{frontmatter}

\section{Introduction}

Recent years have seen several attempts to modify Einstein's theory of gravitation to overcome problems in the context of observational cosmology \cite{Felice:etal:2010}. Furthermore, there is also the continued effort to find a quantum gravitational theory which can explain the emergence of General Relativity (GR) from first principles \cite{Kiefer:2012}.

A common feature of several alternative gravity theories is the appearance of higher-order curvature terms in the action. In particular, recently generalized gravity theories with nonminimal coupling functions that depend on curvature invariants have been proposed \cite{Bertolami:etal:2007,Nojiri:2011}. 

An important question of direct experimental relevance arises in this context: How do test bodies move in such generalized theories? Or, in the words of C.\ Lanczos\footnote{See page 723 in \cite{Lanczos:1927}, translation from the German original by the authors.} -- who was one of early contributors to the so-called ``problem of motion'' \cite{Havas:1986} in Einstein's theory:   
\begin{quote}
``The fact that matter is subject to forces under the influence of an external field appears to be natural and self-evident for an experimental physicist, but is a hard nut for the theoretician to crack.'' 
\end{quote}
There is a fundamental link between conservation laws in gravity theory and equations of motion of {\it test bodies}: the latter are no longer a mere additional postulate. Powerful methods exist that allow to extract the equations of motion of test bodies from the conservation laws of a given theory. In particular, the multipolar methods which go back to the seminal work of Mathisson \cite{Mathisson:1937}, and also underlie the more widely known work of Papapetrou \cite{Papapetrou:1951:3}. They provide the theoretical framework for free fall experiments and also for the GP-B satellite mission \cite{GPBweb,Everitt:etal:2011}.

Here we use the method of multipole expansions to derive the equations of motion for test bodies in a very large class of gravitational theories with a general nonminimal coupling to matter.

\begin{figure}
\begin{center}
\includegraphics[height=6cm]{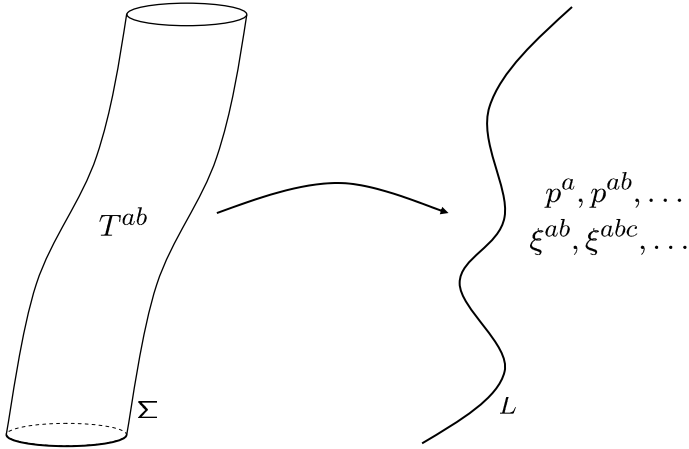}
\end{center}
\caption{General idea behind multipolar approximation schemes: The world-tube $\Sigma$ of a body is replaced by a representative world-line $L$, whereas the original energy-momentum tensor $T^{ab}$ is substituted by a set of multipole moments $p^{a \cdots}$ and $\xi^{ab \cdots}$ along this world-line. Such a multipolar description simplifies the equations of motion. This is achieved by consideration of only a finite set of moments. Different flavors of multipolar approximation schemes exist in the literature, in this work we define the moments  \`{a} la Dixon in \cite{Dixon:1964}.}
\label{fig:1}
\end{figure}

\section{Gravity theories with a general nonminimal coupling}
Let us consider a wide class of models 
\begin{eqnarray} 
L_{\rm tot} =  L_{\rm grav} + F L_{\rm mat}, \label{ansatz_lagrangian_model_2}
\end{eqnarray}
where both the gravitational field Lagrangian $L_{\rm grav} = L_{\rm grav}(g_{ij}, R_{ijk}{}^l)$ and the {\it coupling function} $F = F(g_{ij}, R_{ijk}{}^l)$ can depend arbitrarily on the spacetime metric and the Riemannian curvature tensor.

We assume that the nonminimal model (\ref{ansatz_lagrangian_model_2}) is invariant under spacetime diffeomorphisms (general coordinate transformations). Via the Noether theorem this gives rise to the conservation law for the energy-momentum tensor. Earlier \cite{Koivisto:2006} it was noticed that the nonminimal coupling affects the energy-momentum conservation. In our general framework we find that, despite the complexity of the field equations of this theory, the conservation laws have a remarkably simple form 
\begin{eqnarray}\label{conservation2}
\nabla^i T_{ij} = {\frac 1F} \left(g_{ij} L_{\rm mat} - T_{ij} \right)\nabla^iF.
\end{eqnarray}  
This can be demonstrated by a direct analysis of the field equations \cite{Puetzfeld:Obukhov:2013} for the class of theories in which $F = F(i_1,\dots,i_9)$ depends arbitrarily on the complete set of 9 parity-even curvature invariants. Furthermore, making use of the Lagrange-Noether formalism this result can be extended to the general case when the nonminimal coupling function $F = F(g_{ij}, R_{ijk}{}^l)$ depends {\it arbitrarily on the curvature}. 

Before we turn to the equations of motion of {\it massive} test bodies in the next section, we note that {\it massless} photons still move along null geodesics in the class of theories under consideration. For a nonminimal coupling of the form $F L_{\rm Maxwell}$, with the standard Maxwell field Lagrangian $L_{\rm Maxwell} =  - \,{\frac {\lambda_0}4}F_{ij}F^{ij}$, the coupling function $F$ effectively acts as a dilaton field coupled to the electromagnetic field strength $F_{ij}$. The rays -- paths of light -- are obtained from the geometric optics approximation for the Maxwell field equations. The wave covectors $k_i$ satisfy the Fresnel equation (also known as a dispersion relation), the structure of which is determined by the constitutive tensor $\chi^{ijkl}$ that links the field strength to the tensor of electromagnetic excitations $H^{ij} = {\frac 12}\chi^{ijkl}F_{kl}$, see \cite{Hehl:2003} for further details. In the case of nonminimal coupling under consideration, we find $\chi^{ijkl} = \lambda\,(g^{ik}g^{jl} - g^{jk}g^{il})$, with $\lambda = F\,\lambda_0$ (as usual, $\lambda_0 = \sqrt{\varepsilon_0/\mu_0}$, where $\varepsilon_0$ and $\mu_0$ are the electric and magnetic vacuum constants, respectively). Such a constitutive tensor belongs to the class of electrodynamical models without birefringence \cite{Hehl:Obukhov:2008} with vanishing axion and effective dilaton field $F = F(g_{ij}, R_{ijk}{}^l)$. As a result, the wave covector satisfies 
\begin{equation}
k_ik^i = 0,\qquad k^j\nabla_jk_i = 0,\label{null}
\end{equation} 
which means that light is moving along null geodesics also in the case of nonminimal coupling. 

\begin{figure}
\begin{center}
\includegraphics[height=6cm]{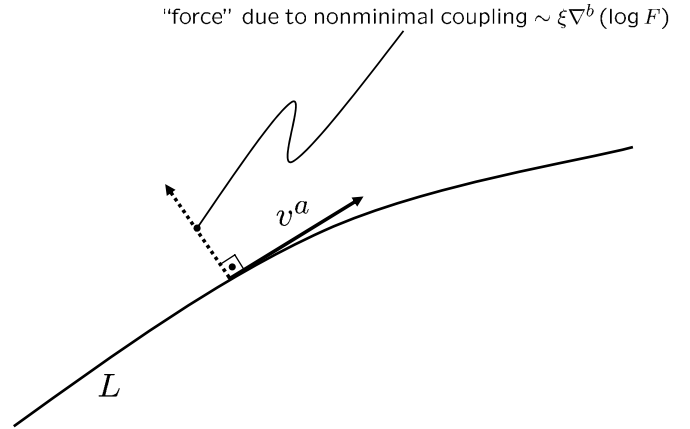}
\end{center}
\caption{The motion of a single-pole (structureless) test body, derived in the lowest order of the multipolar approximation scheme, is non-geodetic, cf.\ eq.\ (\ref{single_eom_2}), but surprisingly simple taking into account the very general class of theories under consideration. The non-geodetic ``force'' is proportional to the gradient of the nonminimal coupling function $F$.}\label{fig:2}
\end{figure}

\section{Equations of motion for test bodies}\label{sec:eom_of_test_bodies}

The covariant multipolar approximation scheme \cite{Dixon:1964} utilizes an expansion technique of Synge \cite{Synge:1960} based on the notion of the world-function. We use this approach to derive the equations of motion for (extended) test bodies from the new energy-momentum conservation law given in (\ref{conservation2}). This can be done to any multipolar order \cite{Puetzfeld:Obukhov:2013}, here we confine our attention to the two lowest orders. See figure \ref{fig:1} for a sketch of the general idea behind multipolar approach in relativistic gravity; note that several different approximation schemes exist in the literature \cite{Mathisson:1937,Papapetrou:1951:3,Dixon:1964,Tulczyjew:1959,Taub:1964,Madore:1969,Dixon:1974}. 

At this stage the consideration can be very general, in particular, it is not important to specify the form of the matter Lagrangian $L_{\rm mat}$. A variety of relativistic Lagrangian models exists in the literature that can be used for the description of matter, e.g., \cite{Taub:1954,Seliger:1968,Schutz:1970,Ray:1972,Obukhov:1987,Obukhov:1989,Brown:1993,Bertolami:etal:2008} to mention but a few. 

We introduce the integrated multipolar moments of the body by:
\begin{eqnarray}
p^{y_1 \cdots y_n y_0}&:=& (-1)^n  \int\limits_{\Sigma(\tau)} \sigma^{y_1} \cdots \sigma^{y_n} g^{y_0}{}_{x_0}  \widetilde{T}^{x_0 x_1} d \Sigma_{x_1}, \label{p_moments_def} \\
\xi^{y_2 \cdots y_{n+1} y_0 y_1}&:=& (-1)^{n}  \int\limits_{\Sigma(\tau)} \sigma^{y_2} \cdots \sigma^{y_{n+1}} \times \nonumber \\  && \times g^{y_0}{}_{x_0} g^{y_1}{}_{x_1} \widetilde{L}_{\rm mat}  
g^{x_0 x_1} w^{x_2} d \Sigma_{x_2}. \label{xi_moments_def}
\end{eqnarray}
Here the tilde marks densities, $\sigma$ denotes Synge's \cite{Synge:1960} world-function and $\sigma^y$ its first covariant derivative; $g^y{}_x$ is the parallel propagator, cf.\ appendix \ref{conventions_app} for more details. 

For spinning test bodies, in the pole-dipole order, we obtain the following set of equations of motion:
\begin{eqnarray}
\frac{D}{d\tau} {\cal P}^a &=& \frac{1}{2} R^a{}_{bcd} v^b{\cal S}^{cd} + \xi^{ab} \nabla_bF + \xi^{cab}\nabla_c\nabla_bF, \label{dipl_eom_1}\\ 
\frac{D}{d\tau}{\cal S}^{ab} &=& 2 v^{[b} {\cal P}^{a]} + 2\xi^{[ab]c}\nabla_cF. \label{dipl_eom_2}
\end{eqnarray}
Here  all quantities are defined -- in a covariant manner -- along the world-line $L$. The velocity is given by $v^{a}:=dx^{a}/d\tau$, where $\tau$ is the proper time. Furthermore, we have introduced the spin of the test body as $s^{ab} := 2 p^{[ab]}$, and denoted the generalized momentum and spin tensors as
\begin{eqnarray}
{\cal P}^a &=& Fp^a + p^{ba}\nabla_b F,\label{Pgen}\\
{\cal S}^{ab} &=& Fs^{ab}.\label{Sgen}
\end{eqnarray}

Equations (\ref{dipl_eom_1}) and (\ref{dipl_eom_2}) represent a generalization of the famous equations of Mathisson \cite{Mathisson:1937} and Papapetrou \cite{Papapetrou:1951:3} for spinning test bodies in GR. Note that (\ref{dipl_eom_1}) and (\ref{dipl_eom_2}) are valid for the {\it whole} class of theories (\ref{ansatz_lagrangian_model_2}) with a general coupling function $F=F(g_{ij},R_{ijk}{}^l)$. In the absence of such a coupling (i.e., when the coupling function becomes the coupling constant, $F=\,$const) our equations reduce to the well-known ones from GR.

It is worthwhile to mention the dual role played by the nonminimal function $F$: On the one hand, it ``rescales'' the ordinary momentum, spin and mass; and on the other hand, its gradients determine the force and torque that act on a particle in addition to the usual gravitational and Mathisson-Papapetrou forces. 

The motion of single-pole test bodies is also non-geodesic. In this case, the geodesic equation, as encountered in GR, is replaced by
\begin{eqnarray}
\frac{D}{d\tau}\left( Fmv^a\right) = \xi^{ab}\nabla_bF.\label{single_eom_1}
\end{eqnarray}
Noticing that $\xi^{ab} = g^{ab}\xi$ with $\xi = \int\limits_{\Sigma(\tau)}\widetilde{L}_{\rm mat}w^{x_2}d\Sigma_{x_2}$, we recast this into
\begin{eqnarray}
m\dot{v}^a = \xi\left(\delta^a_b - v^av_b\right)\nabla^b{\rm log}F.\label{single_eom_2}
\end{eqnarray}
Therefore, a massive particle moves non-geodetically under the action of the ``pressure''-type force (\ref{single_eom_2}) produced by the nonminimal coupling function $F$, cf.\ figure \ref{fig:2}.

The motion of a massive test body (\ref{single_eom_2}) in general depends on the composition of matter through the integrated scalar ``charge'' $\xi$. Experiments testing the universality of free fall put strong limits on such theories (already present day accelerometers reach a sensitivity of $<10^{-12}$ m/s$^2$). In addition, a curvature dependent coupling arises through $F$ -- the specific details are derived from the structure of a particular model.

\section{Conclusions}

In conclusion, we presented covariant equations of motion for extended test bodies for a very large class of non-standard gravity theories with general nonminimal coupling. We found, in particular, that {\it massless} photons move along null geodesics, and thus cannot probe the nonminimal coupling of the type (\ref{ansatz_lagrangian_model_2}). Experimentalists are however encouraged to use our results (\ref{dipl_eom_1})-(\ref{dipl_eom_2}) and (\ref{single_eom_2}) as a universal framework to systematically test the effects of nonminimal coupling by means of spinning, as well as structureless {\it massive} test bodies.

\section*{Acknowledgements}
This work was supported by the Deutsche Forschungsgemeinschaft (DFG) through the grant LA-905/8-1 (D.P.).

\appendix

\section{Conventions \& Symbols}\label{conventions_app}

Our conventions for the Riemann curvature are as follows:
\begin{eqnarray}
&& 2 T^{c_1 \dots c_k}{}_{d_1 \dots d_l ; [ba] } \equiv 2 \nabla_{[a} \nabla_{b]} T^{c_1 \dots c_k}{}_{d_1 \dots d_l} \nonumber \\
& = & \sum^{k}_{i=1} R_{abe}{}^{c_i} T^{c_1 \dots e \dots c_k}{}_{d_1 \dots d_l} \nonumber \\
&& - \sum^{l}_{j=1} R_{abd_j}{}^{e} T^{c_1 \dots c_k}{}_{d_1 \dots e \dots d_l}. \label{curvature_def}
\end{eqnarray}
The Ricci tensor is introduced by $R_{ij} := R_{kij}{}^k$, and the curvature scalar is $R := g^{ij}R_{ij}$. Note that our curvature conventions differ from those in \cite{Synge:1960,Poisson:etal:2011}. The signature of the spacetime metric is assumed to be $(+1,-1,-1,-1)$.

In the derivation of the equations of motion we made use of the bitensor formalism, see, e.g., \cite{Synge:1960,Poisson:etal:2011,DeWitt:Brehme:1960} for introductions and references. In particular, the world-function is defined as an integral $\sigma(x,y) := \frac{1}{2} \epsilon \left( \int\limits_x^y d\tau \right)^2$ over the geodesic curve connecting the spacetime points $x$ and $y$, where $\epsilon = \pm 1$ for timelike/spacelike curves. Indices attached to the world-function always denote covariant derivatives, at the given point, i.e.\ $\sigma_y:= \nabla_y \sigma$, hence we do not make explicit use of the semicolon in case of the world-function. The parallel propagator $g^y{}_x(x,y)$ allows for the parallel transportation of objects along the unique geodesic that links the points $x$ and $y$. For example, given a vector $V^x$ at $x$, the corresponding vector at $y$ is obtained by means of the parallel transport along the geodesic curve as $V^y = g^y{}_x(x,y)V^x$. For more details see, e.g., section 5 in \cite{Poisson:etal:2011}. A compact summary of useful formulas in the context of the bitensor formalism can also be found in the appendices A and B of \cite{Puetzfeld:Obukhov:2013}.

\bibliographystyle{unsrt}

\bibliography{ugbeyondeinstein}

\end{document}